\documentclass[12pt]{article}
\usepackage{amsmath,amssymb}
\begin{document}
\begin{flushright} SU-4252-815
\\
\end{flushright}
\begin{center}
\vskip 3em
{\LARGE Twisted Supersymmetry, Fermion-Boson Mixing and Removal of UV-IR Mixing}
\vskip 2em
{\large B. A. Qureshi$^\dagger$
\footnote[1]{E-mail: bqureshi@phy.syr.edu}
\\[2em]}
\em{\oddsidemargin 0 mm
$^\dagger$Department of Physics, Syracuse University,
Syracuse, NY 13244-1130, USA}\\
\end{center}
\vskip 1em
\begin{abstract}
Exact supersymmetry can be maintained on non-anticommutative
superspace with a twisted coproduct on the supergroup.We show that
the usual exchange statistics for the superfields is not compatible
with the twisted action of the superpoincar\'e group and find a
statistics which is consistent with the twisted coproduct and imply
interesting phenomena such as mixing of fermions and bosons under
particle exchange.We also show that with the new statistics, the
$S$-matrix becomes completely independent of the deformation
parameter.
\end{abstract}
\newpage

\setcounter{footnote}{0}

\section{Introduction}

Recently it has been noted that certain spacetime symmetries which
were thought to be explicitly broken on noncommutative spaces, do
have a well defined action on these noncommutative spaces with a
twisted coproduct. In fact this result was already well formulated
in the theory of quantum groups. For review
see\cite{Dimitrijevic:2004rf,Chaichian:2004za}. A natural
generalization of noncommutative spaces are non-anticommutative
superspaces, where the algebraic relations between superspace
coordinates are deformed. Such spaces have also been shown to arise
in certain limits of string theory\cite{Seiberg}. Just like
noncommutative spaces, these deformed superspaces  break
supersymmetry, but again full supersymmetry can be implemented by
defining a twisted action of superymmetry generators on the product
of two fields\cite{Zupnik,lhl}.

It was shown in \cite{Balachandran:2005eb}, in the context of
Poincar\'e group and Moyal plane, that one can not consistently
impose the usual symmetrization or anti-symmetrization on the tensor
product of fields compatibly with the twisted Poincar\'e symmetry.

In this paper we briefly review the Drinfel'd twist for deformed
superspace and then show that in this case again, one can not impose
the usual statistics. The new statistics gives the novel phenomenon
of fermion-boson mixing under particle exchange. In the end we show
that with the new statistics, the $S$-operator becomes completely
independent of the deformation parameter.

\section{Twisted Superspace and Twisted Supersymmetry}

A simple deformation of superspace consists of the following
superspace algebra

\begin{equation}
\{\theta^\alpha,\,\theta^\beta\,\}=\,C^{\alpha\beta}\nonumber
\end{equation}
\begin{equation}
\{\overline{\theta}^{\dot \alpha},\,\overline{\theta}^{\dot \beta}\,\}=\,\{\overline{\theta}^{\dot \alpha},\,\theta^\beta\}\,=\,[y^\mu,\,y^\nu]\,=\,[y^\mu,\,\theta^\alpha]\,=\,[y^\mu,\,\overline{\theta}^{\dot \alpha}]\,=\,0
\end{equation}

where   $y^\mu=x^\mu+i\theta^\alpha \sigma^{\mu}_{\alpha \dot \alpha}\overline{\theta}^{\dot \alpha}$    is the chiral coordinate.

This deformation of superspace has been considered by Seiberg and others and has been extensively studied over the past years.

The above deformation can be implemented on the usual superfunctions by the star product

\begin{equation}
f\,*\,g=m_\circ \cdot\, \mathcal{F}^{-1}\,(f\,\otimes \,g).
\end{equation}

Where $f$ and $g$ are functions on superspace, $m_\circ$ is the
usual multiplication map of the superspace algebra and

\begin{equation}
\mathcal{F}\,=\,exp(-\frac{1}{2}\,C^{\alpha\beta}\,Q_\alpha\,\otimes
\,Q_\beta).
\end{equation}
$\mathcal{F}$ is called ``the twist element''.

At first sight the above deformation does not seem to preserve supersymmetry, e.g.,
$$
[y^\mu\, ,\,y^\nu]\,=\,0
$$
is not invariant under translations generated by $\overline {Q}$,
hence this space is sometimes said to have $N=1/2$ supersymmetry.
But one can define a new action of the superpoincar\'e generators on
the product of fields ( a deformed Leibniz's rule) so that all the
defining relations are preserved under supersymmetry. The new action
is given in terms of a deformed coproduct of the superpoincar\'e
algebra. The deformed coproduct is given by

\begin{equation}
\Delta_\theta(g)\,=\,\mathcal{F} \,\Delta_\circ(g) \,\mathcal{F}^{-1}
\end{equation}

where $g$ is a superpoincar\'e algebra element and $\Delta_\circ$ is
the usual undeformed coproduct
$$
\Delta_\circ(g)\,=\,1\,\otimes\, g\,+\,g\,\otimes \,1
$$

The action of superalgebra on product of two superfunctions through the deformed coproduct is given by
\begin{equation}
g\triangleright (f\,*\,g)\,=\,m_\circ \,\mathcal{F}^{-1}\,\Delta_\theta(g)\,(f\,\otimes \,g)
\end{equation}

This deformation of the superspace algebra along with the
deformation of the coproduct on superpoincar\'e group( or algebra)
is what is known as  Drinfel'd Twist.
\section{Twisted Statistics}
\vspace{.5cm}

Following\cite{Balachandran:2005eb} , where it was shown that the
usual statistics is incompatible with the twisted coproduct of the
poincar\'e group, we investigate the statistics on deformed
superspace.

Consider the tensor product of two chiral scalar superfields.
Usually we take the tensor product to be symmetric at space-like
separations.

\begin{equation}
\Phi\otimes \Phi (y,\theta\,;\,y',\theta'\,)=\,\Phi\otimes \Phi(y',\theta'\,;\,y,\theta)
\end{equation}

We will show that the above relation on the tensor product of two
scalar superfields is not compatible with the twisted action of the
superpoincar\'e group.

Let us expand the field $\Phi$ into a Fourier expansion
$$
\Phi(y,\theta)\,=\,\int d^4k d^2\kappa\,a_{k,\kappa}\,e^{iky}e^{\kappa \theta}
$$
then the above tensor product can be written as

\begin{equation}
\Phi\otimes \Phi (y,\theta\,;\,y',\theta')\,=\,\int d^4k d^4k' d^2\kappa d^2\kappa'\,a_{k,\kappa}a_{k',\kappa'}\,e^{iky}\,e^{\kappa \alpha}\otimes e^{ik'y'}e^{\kappa' \alpha'}
\end{equation}

Now under a $\overline {Q}$ translation, according to the twisted
coproduct rule, this goes to

\begin{eqnarray}
\lefteqn{\Delta_\theta(g_{\overline Q}(\overline \xi))\,\Phi\otimes \Phi(y,\theta\,;\,y',\theta'){}}\nonumber \\& &{}=\int d^4k\,d^4k'\,d^2\kappa\,d^2\kappa'e^{C^{\alpha\beta}\,k_\mu\,k'_\nu\,\sigma^{\mu}_{\alpha \dot \alpha}\,\sigma^{\nu}_{\beta \dot \beta}\,\overline \xi^{\dot \alpha}\,\overline \xi^{\dot \beta}}\,e^{ik_\mu(y^\mu+2i\theta\sigma^\mu\overline \xi)}\,e^{\kappa \theta}\otimes e^{ik'_\nu(y'^\nu+2i\theta'\sigma^\nu\overline \xi)}\,e^{\kappa'\theta'}
\end{eqnarray}

Now under the permutation of the two fields we get

\begin{eqnarray}
\lefteqn{\sigma\Delta_\theta(g_{\overline Q}(\overline \xi))\,\Phi\otimes \Phi(y,\theta\,;\,y',\theta'){}}\nonumber \\& &{}=\int d^4k\,d^4k'\,d^2\kappa\,d^2\kappa'e^{C^{\alpha\beta}\,k_\mu\,k'_\nu\,\sigma^{\mu}_{\alpha \dot \alpha}\,\sigma^{\nu}_{\beta \dot \beta}\,\overline \xi^{\dot \alpha}\,\overline \xi^{\dot \beta}}\,e^{ik'_\nu(y'^\nu+2i\theta'\sigma^\nu\overline \xi)}\,e^{\kappa' \theta'}\otimes e^{ik_\mu(y^\mu+2i\theta\sigma^\mu\overline \xi)}\,e^{\kappa\theta} \label{eq:9}
\end{eqnarray}

where $\sigma$ is the permutation which interchanges the position of the two fields in the tensor product.

On the other hand if we first apply the permutation and then perform
the super transformation, using $\overline Q$ we get

\begin{eqnarray}
\lefteqn{\Delta_\theta(g_{\overline Q}(\overline \xi))\,\sigma\,\Phi\otimes \Phi(y,\theta\,;\,y',\theta'){}}\nonumber \\& &{}=\int d^4k\,d^4k'\,d^2\kappa\,d^2\kappa'e^{C^{\alpha\beta}\,k_\mu\,k'_\nu\,\sigma^{\nu}_{\alpha \dot \alpha}\,\sigma^{\mu}_{\beta \dot \beta}\,\overline \xi^{\dot \alpha}\,\overline \xi^{\dot \beta}}\,e^{ik'_\nu(y'^\nu+2i\theta'\sigma^\nu\overline \xi)}\,e^{\kappa' \theta'}\otimes e^{ik_\mu(y^\mu+2i\theta\sigma^\mu\overline \xi)}\,e^{\kappa\theta} {}\nonumber \\& &{}=\int d^4k\,d^4k'\,d^2\kappa\,d^2\kappa'e^{-\,C^{\alpha\beta}\,k_\mu\,k'_\nu\,\sigma^{\mu}_{\alpha \dot \alpha}\,\sigma^{\nu}_{\beta \dot \beta}\,\overline \xi^{\dot \alpha}\,\overline \xi^{\dot \beta}}\,e^{ik'_\nu(y'^\nu+2i\theta'\sigma^\nu\overline \xi)}\,e^{\kappa' \theta'}\otimes e^{ik_\mu(y^\mu+2i\theta\sigma^\mu\overline \xi)}\,e^{\kappa\theta} \label{eq:10}
\end{eqnarray}

where in the last line we have used the fact that $C^{\alpha \beta}$ is symmetric in $\alpha,\beta$.

From (\ref{eq:9}) and (\ref{eq:10}) we have that

\begin{equation}
\sigma\,\Delta(g)\,\Phi\otimes \Phi\,\not =\,\Delta(g)\,\sigma\,\Phi\otimes \Phi.
\end{equation}

This means that the usual statistics is incompatible with the
twisted action of superymmetry, and we can not impose symmetric
statistics on the tensor product of fields .

But we can impose

\begin{equation}
\Phi\otimes \Phi (y,\theta\,;\,y',\theta'\,)=\,\mathcal{F}^{-2}\,\Phi\otimes \Phi(y',\theta'\,;\,y,\theta) \label{eq:12}
\end{equation}

consistently with the twisted action of superpoincar\'e group. These
relations reduce to standard symmetrization when the deformation
parameter $C^{\alpha\beta}$ goes to zero.

One can check that with this deformed permutation, which we will call $\sigma_\theta$, we have

\begin{eqnarray}
\lefteqn{\sigma_\theta\Delta_\theta(g)\,\Phi\otimes \Phi(y,\theta\,;\,y',\theta'){}}\nonumber \\& &{}=\int d^4k\,d^4k'\,d^2\kappa\,d^2\kappa'e^{-\,C^{\alpha\beta}\,k_\mu\,k'_\nu\,\sigma^{\mu}_{\alpha \dot \alpha}\,\sigma^{\nu}_{\beta \dot \beta}\overline \xi^{\dot \alpha}\,\overline \xi^{\dot \beta}}\,e^{C^{\alpha\beta}\kappa_\alpha\kappa_\beta}\,e^{ik'_\nu(y'^\nu+2i\theta'\sigma^\nu\overline \xi)}\,e^{\kappa' \theta'}\otimes e^{ik_\mu(y^\mu+2i\theta\sigma^\mu\overline \xi)}\,e^{\kappa\theta}\nonumber
\end{eqnarray}

and exactly the same expression for
$\Delta_\theta(g)\,\sigma_\theta\,\Phi\otimes\Phi$. Same statements
can be proved for the transformations generated by other elements of
the superpoincare\'e algebra.
\section{Fermion-Boson Mixing}

\vspace{.5cm}

The above relations on the tensor product of two chiral fields imply interesting statistics for the component fields. The new statistics mixes the bosons with fermions under an exchange of particle. Expnding the both sides of (\ref{eq:12})  into component fields, we find the relations

\begin{equation}
A(y)A(y')\,=\,A(y')A(y)\,-\,C^{\alpha\beta}\,\psi_\alpha(y')\psi_\beta(y)\,+\,\frac{1}{2}\,C^{\alpha\beta}C_{\alpha\beta}\,F(y')F(y) \nonumber
\end{equation}

\begin{equation}
\psi_\alpha(y)\psi_\beta(y')\,=\,-\,\psi_\beta(y')\psi_\alpha(y)\,+\,C_{\alpha\beta}F(y')F(y) \nonumber
\end{equation}

\begin{equation}
F(y)F(y')\,=\,F(y')F(y)
\end{equation}

\section{Removal of UVIR mixing}

As an example of the use of twisted statistics, we show  that the
$S$-matrix, in this formalism becomes completely independent of the
deformation parameter $C^{\alpha\beta}$

Let's take $\Phi$ to be a free chiral scalar superfield, and
consider an interaction Hamiltonian of the form

\begin{equation}
H_I(x_o)\,=\,\lambda\,\int d^3x\,d^2\theta\,d^2\overline \theta\,\delta(\overline \theta)\,\Phi\ast\Phi\ast\Phi
\end{equation}

The $S$-matrix is

\begin{eqnarray}
\lefteqn{ S_\theta = T \, \exp\left(\ -\ \int d x_0 \, H_I(x_0)\right) {}} \nonumber \\
& & {} = T \, \exp\left(\ -\ \int d^{4} x\,d^2\theta\,d^2\overline \theta\,\delta(\overline \theta)\,\Phi(y,\theta)\ast\Phi(y,\theta)\ast\Phi(y,\theta)
\right) \nonumber
\end{eqnarray}

here $\theta$ in $S_\theta$ just reminds us that our theory is on deformed superspace.

The fields $\Phi$ obey the twisted statistics

$$
\Phi(y,\theta)\,\Phi(y',\theta')=e^{C^{\alpha\beta}\partial_{\theta\alpha}\partial_{\theta'\beta}}\ \Phi(y',\theta')\,\Phi(y,\theta)
$$

We can take care of this statictics by writing the fields as

\begin{equation}
\Phi\,=\,\Phi_o\,e^{-\frac{1}{2}C^{\alpha\beta}\overleftarrow \partial_\alpha\,\overrightarrow \partial_\beta} \label{eq:15}
\end{equation}

where $\Phi_o$ has the usual statistics, the differential to the
left acts only on the field and the differential to the right acts
on every thing to the right of it.

For example, consider the product of the fields (from now on we will
suppress the y dependence of the fields)

\begin{eqnarray}
\lefteqn{\Phi(\theta)\Phi(\theta')=\,\Phi_o(\theta)\,e^{-\frac{1}{2}C^{\alpha\beta}\overleftarrow \partial_\alpha\,\overrightarrow \partial_\beta}\,\Phi_o(\theta')\,e^{-\frac{1}{2}C^{\alpha\beta}\overleftarrow \partial_\alpha\,\overrightarrow \partial_\beta}{}}\nonumber \\
& &{}= e^{-\frac{1}{2}\,C^{\alpha\beta}\,\partial_{\alpha\theta}\, \partial_{\beta\theta'}}\,\Phi_o(\theta)\Phi_o(\theta')\,e^{-\frac{1}{2}C^{\alpha\beta}(\overleftarrow \partial_{\alpha\theta}+\overleftarrow\partial_{\beta\theta'})\,\overrightarrow \partial_\beta} \nonumber \\
& & {}= e^{+\frac{1}{2}\,C^{\alpha\beta}\,\partial_{\alpha\theta'}\,\partial_{\beta\theta}}\,\Phi_o(\theta')\Phi_o(\theta)\,e^{-\frac{1}{2}C^{\alpha\beta}\,(\overleftarrow\partial_{\alpha\theta}+\overleftarrow\partial_{\beta\theta'})\,\overrightarrow \partial_\beta} \nonumber \\
& & {}=e^{+C^{\alpha\beta}\,\partial_{\alpha\theta'}\,\partial_{\beta\theta}}\,\Phi(\theta')\Phi(\theta)
\end{eqnarray}

showing that the field $\Phi$ has the correct statistics.

Now let us consider the $\it{O}(\lambda^2)$ in the $S$-matrix.

\begin{equation}
S^{(2)}_\theta\,\sim\,T\ \int\,d^4x\,d^4x'\,d^2\theta\,d^2\theta'\,d^2\overline \theta\,d^2\overline \theta'\,\delta(\overline \theta)\,\delta(\overline \theta')\,\Phi\ast\Phi\ast\Phi(\theta)\ \Phi\ast\Phi\ast\Phi(\theta')
\end{equation}

Consider

$$
\Phi\ast\Phi\ast\Phi(\theta)=\Phi\,e^{\frac{1}{2}C^{\alpha\beta}\,\overleftarrow\partial_\alpha\,\overrightarrow \partial_\beta}\Phi\,e^{\frac{1}{2}C^{\alpha\beta}\,\overleftarrow\partial_\alpha\,\overrightarrow \partial_\beta}\Phi(\theta)
$$

Substituting the expression for $\Phi$ from (\ref{eq:15}), we have

\begin{equation}
\Phi\ast\Phi\ast\Phi(\theta)\,=\,\Phi_o\,\Phi_o\,\Phi_o(\theta)\,e^{-\,\frac{1}{2}C^{\alpha\beta}\,\overleftarrow\partial_\alpha\,\overrightarrow \partial_\beta}
\end{equation}

Then the relevant integral becomes

\begin{eqnarray}
\lefteqn{\int\,d^2\theta\,d^2\theta'\,\Phi\ast\Phi\ast\Phi(\theta)\,\Phi\ast\Phi\ast\Phi(\theta'){}}\nonumber \\
& & {} =\,\int\,d^2\theta\,d^2\theta'\,e^{-\,\frac{1}{2}C^{\alpha}{\beta}\,\partial_{\alpha\theta}\,\partial{\beta\theta'}}\,\Phi_o\,\Phi_o\,\Phi_o(\theta)\,
\Phi_o\,\Phi_o\,\Phi_o(\theta')\,e^{-\,\frac{1}{2}C^{\alpha\beta}\,(\overleftarrow\partial_{\alpha\theta}\,+\,\overleftarrow\partial_{\alpha\theta'})\,\overrightarrow \partial_\beta} \\
& & {}=\,\int\,d^2\theta\,d^2\theta'\,\Phi_o\,\Phi_o\,\Phi_o(\theta)\,
\Phi_o\,\Phi_o\,\Phi_o(\theta')
\end{eqnarray}

where in the last line we have used
$$
\int d\theta\,\partial_\theta\,f(\theta)=0
$$

But then we have removed all  dependence on $C^{\alpha\beta}$ from the $\it{O}(\lambda^2)$ term and reduced it to the undeformed form i.e.,
\begin{equation}
S^{(2)}_\theta\,=\,S^{(2)}_o
\end{equation}

This proof generalizes straight-forwardly to higher orders in
$\lambda$ and the proof to first order is even simpler. Hence we
have proved, to all orders in perturbation theory
$$
S_\theta\,=\,S_o
$$

\vspace{.5cm}

{\bf Acknowledgments}

\vspace{.5cm}

The author would like to thank Prof. A. P. Balachandran for suggesting this problem,and also thank Prof. A. P. Balachandran and A. Pinzul for many helpful discussions.. The work was
supported by DOE under grant number DE-FG02-85ER40231 and by NSF
under contract number INT9908763.


\begin{thebibliography}{12}




\bibitem{Dimitrijevic:2004rf}
  M.~Dimitrijevic and J.~Wess,
  ``Deformed bialgebra of diffeomorphisms,''
  arXiv:hep-th/0411224;
  P.~Aschieri, C.~Blohmann, M.~Dimitrijevic, F.~Meyer, P.~Schupp and J.~Wess,
  ``A gravity theory on noncommutative spaces,''
  arXiv:hep-th/0504183.

\bibitem{Chaichian:2004za}
  M.~Chaichian, P.~P.~Kulish, K.~Nishijima and A.~Tureanu,
  ``On a Lorentz-invariant interpretation of noncommutative space-time and its
  implications on noncommutative QFT,''
  Phys.\ Lett.\ B {\bf 604}, 98 (2004)
  [arXiv:hep-th/0408069];
  M.~Chaichian, P.~Presnajder and A.~Tureanu,
  ``New concept of relativistic invariance in NC space-time: Twisted  Poincare
  symmetry and its implications,''
  Phys.\ Rev.\ Lett.\  {\bf 94}, 151602 (2005)
  [arXiv:hep-th/0409096].


\bibitem{Grosse:2001mar}
 H. Grosse, J. Madore, H. Steinacker,
 ``Field theory on the $q$-deformed Fuzzy Sphere II: Quantization'',
 J.Geom.Phys. 43(2002) 205-240
 [arXiv:hep-th/0103164].

\bibitem{Seiberg}
Nathan Seiberg,
``Noncommutative Superspace, N=1/2 Supersymmetry, Field Theory and String Theory'',
JHEP 0306 (2003) 010
[arXiv:hep-th/0305248].

\bibitem{Zupnik}
B. M. Zupnik,
``Twist-Deformed Supersymmetry in Non-Anticommutative Superspaces'',
Phys. Lett. B627 (2005) 208-216
[arXiv:hep-th/0506043]

\bibitem{lhl}
Mathias lhl, Christian Saemann,
``Drinfeld-Twisted Supersymmetry and Non-Anticommuttive Superspace'',
JHEP 0601 (2006) 065
[arXiv:hep-th/0506057]


\bibitem{Balachandran:2005eb}
  A.~P.~Balachandran, G.~Mangano, A.~Pinzul and S.~Vaidya,
  ``Spin and statistics on the Groenwold-Moyal plane: Pauli-forbidden levels
  and transitions,''
  [arXiv:hep-th/0508002].






\bibitem{Balachandran:2005aug}
A. P. Balachandran, A. Pinzul, B. Qureshi,
``UV-IR Mixing in Noncommutative Plane'',
[arXiv:hep-th/0508151].

\end{thebibliography}
\end{document}